\documentclass[aps,prd,reprint,twocolumn,superscriptaddress,preprintnumbers,nofootinbib]{revtex4-1}

\usepackage{amsthm}
\usepackage{amsmath}
\usepackage{graphicx}
\usepackage{slashed}
\usepackage{amssymb}
\usepackage{float}
\usepackage[utf8]{inputenc}
\usepackage[T1]{fontenc}
\usepackage[colorlinks=True, citecolor=blue, urlcolor=blue, linkcolor=blue]{hyperref}
\usepackage{bbm}

\newcommand*\diff{\mathop{}\!\mathrm{d}}

\newcommand{\nn}{\nonumber}

\newcommand{\be}{\begin{eqnarray}}
\newcommand{\ee}{\end{eqnarray}}

\newcommand{\ml}{\mathcal}
\newcommand{\bs}{\boldsymbol}
\newcommand{\Tr}{\mathrm{Tr}}

\begin{document}

\title{Generalized Gluon Distribution for Quarkonium Dynamics in Strongly Coupled $\mathcal{N}=4$ Yang-Mills Theory}

\author{Govert Nijs}
\email{govert.nijs@cern.ch}
\affiliation{Center for Theoretical Physics, Massachusetts Institute of Technology, Cambridge, MA 02139, USA}
\affiliation{Theoretical Physics Department, CERN, CH-1211 Gen\`eve 23, Switzerland}

\author{Bruno Scheihing-Hitschfeld}
\email{bscheihi@mit.edu}
\affiliation{Center for Theoretical Physics, Massachusetts Institute of Technology, Cambridge, MA 02139, USA}

\author{Xiaojun Yao}
\email{xjyao@uw.edu}
\affiliation{InQubator for Quantum Simulation, University of Washington, Seattle, WA 98195, USA}

\date{\today}
\preprint{CERN-TH-2023-170, MIT-CTP/5629, IQuS@UW-21-065}
\begin{abstract}
We study the generalized gluon distribution that governs the dynamics of quarkonium inside a non-Abelian thermal plasma characterizing its dissociation and recombination rates. This gluon distribution can be written in terms of a correlation function of two chromoelectric fields connected by an adjoint Wilson line. We formulate and calculate this object in $\mathcal{N}=4$ supersymmetric Yang-Mills theory at strong coupling using the AdS/CFT correspondence, allowing for a nonzero center-of-mass velocity $v$ of the heavy quark pair relative to the medium. The effect of a moving medium on the dynamics of the heavy quark pair is described by the simple substitution $T \to \sqrt{\gamma} \, T$ in agreement with previous calculations of other observables at strong coupling, where $T$ is the temperature of the plasma in its rest frame, and $\gamma = (1 - v^2)^{-1/2}$ is the Lorentz boost factor. Such a velocity dependence can be important when the quarkonium momentum is larger than its mass. Contrary to general expectations for open quantum systems weakly coupled with large thermal environments, the contributions to the transition rates that are usually thought of as the leading ones in Markovian descriptions vanish in this strongly coupled plasma. This calls for new theoretical developments to assess the effects of strongly coupled non-Abelian plasmas on in-medium quarkonium dynamics. Finally, we compare our results with those from weakly coupled QCD, and find that the QCD result moves toward the $\mathcal{N}=4$ strongly coupled result as the coupling constant is increased within the regime of applicability of perturbation theory.
This behavior makes it even more pressing to develop a non-Markovian description of quarkonium in-medium dynamics.

\end{abstract}

\maketitle

Strongly coupled systems, such as superconductors, topological insulators, cold atoms in optical lattices and neutron stars, usually exhibit complex behavior.
Historically, studying them has led to many breakthroughs in our understanding of matter.
One particular example in high energy nuclear physics is the quark-gluon plasma (QGP) created in relativistic heavy ion collisions (HICs). In these experiments, two heavy nuclei are accelerated to almost the speed of light and then collide. Shortly after the collision, a hot and dense QGP is created that only lasts for a tiny fraction of a second ($10^{-22}$ s)\@. The QGP's short lifetime makes it very challenging to measure its properties directly, and so indirect probes have been primarily used.
The microscopic nature of the QGP at different energy scales is studied by combining experimental measurements, phenomenological studies and theoretical calculations at weak and strong coupling.

A useful probe of the QGP involves quarkonium~\cite{Mocsy:2013syh,Chapon:2020heu}\@, a bound state of a heavy quark-antiquark ($Q\bar{Q}$) pair. Different quarkonium species have hierarchically ordered binding energies and thus can probe the QGP at multiple scales. For a long time, it was believed that the suppression of quarkonium production in HICs probes the Debye screening of (the real part of) the $Q\bar{Q}$ potential~\cite{Matsui:1986dk,Karsch:1987pv}\@. However, systematic studies using thermal field theory showed that in addition to the Debye screening, the in-medium $Q\bar{Q}$ potential also develops a thermal imaginary part~\footnote{Whether or not the dissociation rate is the expectation value of the imaginary part of the potential depends on the definition of the potential, i.e., at which scale each relevant process happens.}, which is a reflection of quarkonium dissociation~\cite{Laine:2006ns,Beraudo:2007ky}\@. When the QGP temperature is low enough that a particular $Q\bar{Q}$ bound state can exist, the inverse process of dissociation, i.e., regeneration, also occurs and plays a crucial role in charmonium production~\cite{Thews:2000rj,Andronic:2003zv,Andronic:2007bi}\@. Many phenomenological studies of quarkonium suppression have shown that the dynamical processes of dissociation and regeneration are, if not more important than, as important as the Debye screening~\cite{Krouppa:2015yoa,Du:2017qkv,Yao:2020xzw,Brambilla:2022ynh,Song:2023zma}\@.

\begin{figure}[t]
\centering
\includegraphics[width=2.5in]{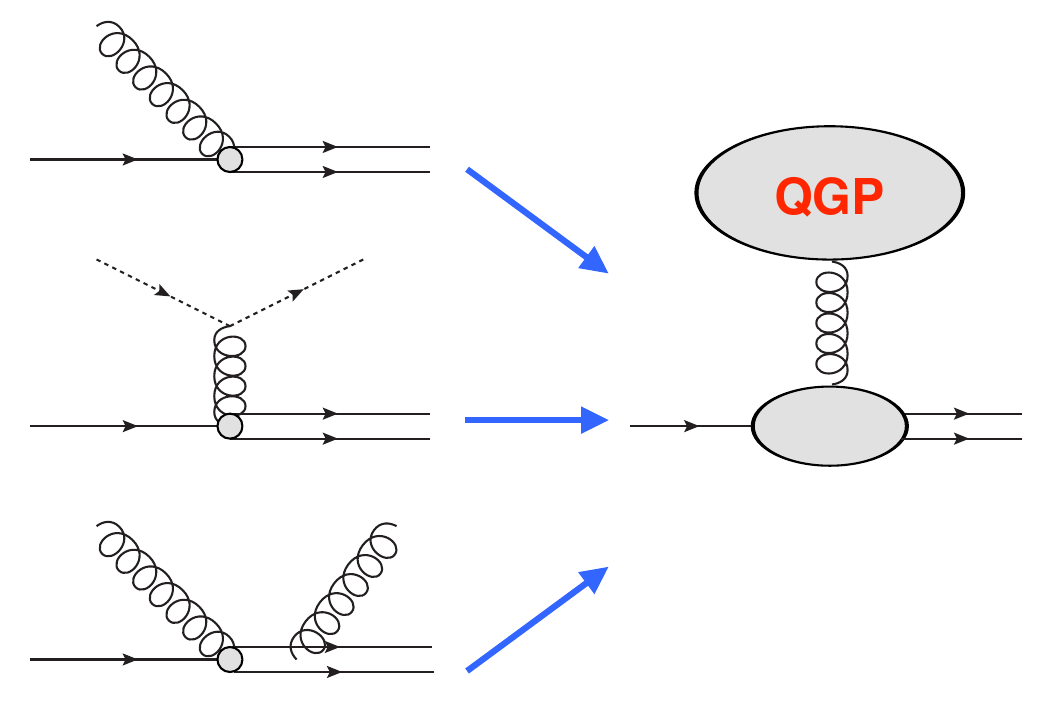}
\caption{A few perturbative Feynman diagrams for quarkonium dynamics (left) and its nonperturbative generalization (right) at leading order (dipole) in the multipole expansion. The single (double) solid line indicates a $Q\bar{Q}$ bound (unbound) state and the dotted line represents a light quark. The effective operator on the right is a chromoelectric field dressed with a timelike adjoint Wilson line.}
\label{fig:f_g}
\end{figure}

The understanding of dynamical processes for quarkonium can be dated back to the early work by Peskin and Bhanot~\cite{Peskin:1979va,Bhanot:1979vb}, where they studied perturbatively the scattering process $g+(Q\bar{Q})_b \leftrightarrow Q+\bar{Q}$ (the subscript $b$ indicates a bound state), as shown in Fig.~\ref{fig:f_g}\@, in which the gluon is on shell. By convoluting the scattering amplitude squared with the Bose-Einstein distribution $n_B$ for the gluon, one can obtain the dissociation rate~\cite{Grandchamp:2001pf,Wong:2004zr,Rapp:2009my,Brambilla:2011sg} and the regeneration rate~\cite{Wong:2004zr,Yao:2017fuc} if the QGP were a free gas of quarks and gluons. These studies have been generalized to the case of a weakly interacting gas in which the gluon mediating the $t$-channel $2\leftrightarrow3$ scattering processes ($q/g+(Q\bar{Q})_b \leftrightarrow q/g+Q+\bar{Q}$) is virtual \footnote{Other non-$t$-channel processes such as those shown in Fig.~\ref{fig:f_g} also contribute, which is a requirement of gauge invariance~\cite{Yao:2018sgn}. A complete set of diagrams at $g^4$ can be found in~\cite{Yao:2018sgn,Binder:2021otw}\@.}~\cite{Brambilla:2013dpa,Yao:2018sgn}\@. 
However, it is well known that at temperatures around $\Lambda_{\rm QCD}$ the QGP is a strongly coupled fluid. This is the regime where most regeneration occurs and the binding energy cannot be neglected. 
Therefore, it is important to find the nonperturbative generalization of the Peskin-Bhanot and related higher order processes.

With recent developments combining potential nonrelativistic QCD and open quantum systems~\cite{Akamatsu:2011se,Akamatsu:2014qsa,Katz:2015qja,Brambilla:2016wgg,Brambilla:2017zei,Blaizot:2017ypk,Kajimoto:2017rel,Blaizot:2018oev,Yao:2018nmy,Akamatsu:2018xim,Miura:2019ssi,Sharma:2019xum,Yao:2020eqy,Akamatsu:2021vsh,Brambilla:2021wkt,Miura:2022arv,Brambilla:2022ynh,Xie:2022tzs,alalawi2023impact} (see recent reviews~\cite{Rothkopf:2019ipj,Akamatsu:2020ypb,Sharma:2021vvu,Yao:2021lus}), a factorization formula was constructed for the dissociation and recombination of small-size quarkonium states~\cite{Yao:2020eqy}\@. At linear order in the multipole expansion, the dissociation and recombination rates are factorized into a nonrelativistic part that only involves the wavefunctions of the $Q\bar{Q}$ pair, which can be obtained from solving Schr\"odinger equations, and a generalized gluon distribution (GGD), shown in Fig.~\ref{fig:f_g}\@, which is the effective distribution of quasi-gluons from the medium that the $Q\bar{Q}$ pair absorbs or radiates. This GGD has only been studied perturbatively so far~\cite{Binder:2021otw}\@. In this letter, we report the first nonperturbative study of it that uses the AdS/CFT correspondence by extending the framework in~\cite{Nijs:2023dks} to the case of a $Q\bar{Q}$ pair moving in a thermal plasma, and compare with its weakly coupled counterpart in QCD\@. Surprisingly, our findings suggest that a small-size $Q\bar{Q}$ pair weakly interacting with a strongly coupled plasma is an exception to the general expectation~\cite{Breuer:2002pc} that the dynamics of an open quantum system weakly coupled with a large thermal environment can be described by Markovian processes, and therefore, that the existing transport formalisms need to be generalized to include this regime.

{\it Generalized gluon distribution.} We first review the factorization formula for quarkonium dissociation and recombination and the relevant GGD, valid in the quantum optical regime $M\gg Mv_{\rm rel}\gg Mv^2_{\rm rel},T$ where $v_{\rm rel}$ is the relative velocity between the heavy quark pair and $T$ the plasma temperature.
The number density $n_b$ of a quarkonium state with quantum numbers $b$ evolves in time according to~\cite{Yao:2018nmy}
\begin{align}
\label{eqn:rate}
\frac{{\rm d} n_b(t,{\bf x})}{{\rm d} t} = -\Gamma\, n_b(t,{\bf x}) + F(t,{\bf x}) \,,
\end{align}
where $\Gamma$ is the dissociation rate and $F$ denotes the contribution of quarkonium (re)combination:
\begin{align}
\label{eqn:disso}
\Gamma &=  \int \! \frac{{\rm d}^3p_{{\rm rel}}}{(2\pi)^3} 
| \langle \psi_b | {\bs r} | \Psi_{{\bs p}_{\rm rel}} \rangle |^2 [g^{++}_{\rm adj}]^{>}(-\Delta E) \\
F &=   \int \! \frac{{\rm d}^3p_{{\rm cm}}}{(2\pi)^3}  \frac{{\rm d}^3p_{{\rm rel}}}{(2\pi)^3} 
| \langle \psi_b | {\bs r} | \Psi_{{\bs p}_{\rm rel}} \rangle |^2
[g^{--}_{\rm adj}]^{>}(\Delta E) f_{Q\bar{Q}}
\nonumber \,.
\end{align}
Here $\Delta E \equiv |E_b| + \frac{p^2_{\rm rel}}{M}$ is the energy transferred to or away from the $Q\bar{Q}$ pair, $\langle \psi_b | {\bs r} | \Psi_{{\bs p}_{\rm rel}} \rangle$ is the dipole transition amplitude between a bound $\psi_b$ and a scattering $\Psi_{{\bs p}_{\rm rel}}$ state, and $f_{Q\bar{Q}}$ is the two-particle phase space distribution. Details can be found in~\cite{Yao:2018nmy}\@.

The GGD for dissociation is defined in terms of a chromoelectric field correlator
\begin{align}
[g_{\rm adj}^{++}]^>(\omega) \equiv \frac{g^2 T_F }{3 N_c} \int\frac{\diff t}{2\pi} e^{i\omega t} \langle E_i^a(t) \ml{W}^{ab}(t,0) E_i^b(0) \rangle_T \,,
\end{align}
where $\ml{W}^{ab}(t,0)$ denotes a timelike adjoint Wilson line and $\langle O \rangle_T = \Tr_\ml{H}(O e^{-H/T})/\ml{Z}$, where $\ml{Z}=\Tr_\ml{H}(e^{-H/T})$, 
and $H$ is the QGP Hamiltonian in the absence of any external sources. The GGD $[g_{\rm adj}^{--}]^>$ for recombination can be related to $[g_{\rm adj}^{++}]^>$ via a generalized Kubo-Martin-Schwinger (KMS) relation
\be \label{eqn:kms}
[g_{\rm adj}^{++}]^>(\omega) = e^{\omega/T} [g_{\rm adj}^{--}]^>(-\omega) \,,
\ee
which is necessary for the system to reach detailed balance between dissociation and recombination~\cite{Binder:2021otw}\@. For a free gluon gas, $[g_{\rm adj}^{++}]^>(-\Delta E) \propto \Delta E^3 n_B(\Delta E)$, $[g_{\rm adj}^{--}]^>(\Delta E) \propto \Delta E^3 ( 1 + n_B(\Delta E))$ and Eq.~\eqref{eqn:disso} reproduces the Peskin-Bhanot result.

Furthermore, the GGDs at zero frequency $g_{\rm adj}^{\pm\pm}(\omega=0)$ govern the Lindblad equation for quarkonium dynamics in the quantum Brownian motion regime $M\gg Mv_{\rm rel} \gg T \gg Mv_{\rm rel}^2$~\cite{Brambilla:2017zei}, where the binding energy effect is suppressed.

{\it Chromoelectric field correlator from a Wilson loop.} For our AdS/CFT calculation, we express the chromoelectric field correlator defining the GGD in terms of variations of a Wilson loop. As an intermediate step, we introduce a time-ordered chromoelectric correlator 
\begin{align} \label{eq:correlator-def}
[g_{\rm adj}^{\ml{T}}]_{ij}(t) \equiv \frac{g^2 T_F }{3 N_c} \langle \hat{\ml{T}} E_i^a(t) \ml{W}^{ab}(t,0) E_j^b(0) \rangle_T \,,
\end{align}
where $\hat{\ml{T}}$ is the time-ordering symbol. 

We consider a closed path $\mathcal{C}_h \subset {\rm Mink}_4$ (4-dimensional Minkowski spacetime) parametrized by $s\mapsto x_h^{\mu}(s) = x^{\mu}_0(s) + h^{\mu}(s)$, where $x^{\mu}_0(s)$ goes along the time direction from $x^0_0=-\ml{T}/2$ to $\ml{T}/2$ and then backtracks to $-\ml{T}/2$. $h^{\mu}(s)$ is a local deformation of this path. A fundamental Wilson loop along $\mathcal{C}_h$ is denoted as $W[\mathcal{C}_h]$\@. First we note that $W[\mathcal{C}_{h=0}] = 1$\@. For $h\neq 0$, $\mathcal{C}_h$ is not made up of two antiparallel straight lines, but if they are still coincident and locally antiparallel at every point, then one still has $W[\mathcal{C}_{h}] = 1$\@. Therefore, nontrivial operator insertions are generated by taking $h^\mu(s)$ to be ``antisymmetric'' on opposite sides of the contour. Technical details are in~\cite{Nijs:2023dks}\@. 
Restricting the variations to be antisymmetric in this sense, we find
\begin{equation}
    -12 [g_{\rm adj}^\ml{T}]_{ij}(t) 
    =
\left. \frac{\delta^2 \langle \hat{\mathcal{T}} W[\mathcal{C}_h] \rangle_T}{\delta h^{i}(t) \delta h^{j}(0)}  \right|_{ h = 0} \label{eq:correlator}
\end{equation}
where we have assumed $0 < |t| < \mathcal{T}/2$.
As desired, we have obtained a timelike adjoint Wilson line $\mathcal{W}^{ab}(t,0)$ from the variation of a fundamental Wilson loop, with electric field operators inserted at its endpoints.

{\it Wilson loops in AdS/CFT\@.}
We will use the holographic correspondence to calculate the Wilson loop in Eq.~\eqref{eq:correlator} in the strong coupling limit.
While this calculation cannot be carried out in QCD, since it does not have a known gravitational dual, it can be carried out in $\mathcal{N} = 4$ supersymmetric Yang-Mills theory ($\mathcal{N}=4$ SYM), which has a well-established gravitational dual description in terms of an (asymptotically) ${\rm AdS}_5 \times S_5$ spacetime~\cite{Maldacena:1997re,casalderrey2014gauge}\@.

There is a well-known prescription~\cite{Maldacena:1998im} to calculate the expectation value of the generalized Wilson loop
\begin{equation} \label{eq:Wloop-S}
    W_S[\mathcal{C}, \hat{n}] = \frac{1}{N_c} {\rm Tr} P e^{ i g \oint_{\mathcal{C}} {\rm d}s \left[ A_\mu(x) \dot{x}^{\mu} + \sqrt{\dot{x}^2} \hat{n} \cdot \phi(x) \right] } \, ,
\end{equation}
where $P$ denotes path ordering, $x^\mu = x^{\mu}(s)$ is the position in ${\rm Mink}_4$ that parametrizes the path $\mathcal{C}$, $\hat{n} = \hat{n}(s)$ describes a path in $S_5$, and $A_\mu, \phi$ are the gauge and scalar fields of $\mathcal{N}=4$ SYM. The expectation value of this loop is given by
\begin{equation}
    \langle W_S[\mathcal{C}, \hat{n}] \rangle = e^{i \left( S_{\rm NG}[\Sigma_{[\mathcal{C},\hat{n}]}] - S_0[\mathcal{C}] \right)} \, ,
\end{equation}
where $\Sigma_{[\mathcal{C},\hat{n}]}$ is the two-dimensional surface with boundary conditions set by $\ml{C}$ and $\hat{n}$, parametrized by $X^{\mu}(\tau, \sigma) = (t(\tau,\sigma), {\bs x}(\tau, \sigma), z(\tau, \sigma), \hat{n}(\tau,\sigma) ) \in {\rm AdS}_5 \times S_5$ that extremizes the Nambu-Goto (NG) action
\begin{equation}
    S_{\rm NG} = -\frac{1}{2\pi \alpha'} \int {\rm d}\tau {\rm d}\sigma  \sqrt{ - \det \left( g_{\mu \nu} \partial_\alpha X^\mu \partial_\beta X^\nu \right) } \,,
\end{equation}
where $2\pi \alpha'$ is the inverse string tension.
Locally, near the boundary $z = 0$, we can choose the coordinates $(\tau,\sigma)$ to be $(s,z)$, where $s$ is the parameter that defines the contour $\mathcal{C}$\@. In these coordinates, the boundary conditions are given by $t(s, 0) = x^0(s)$, ${\bs x}^i(s,0) = x^i(s)$, $\hat{n}(s,0) = \hat{n}(s)$\@.
Finally, $S_0[\mathcal{C}]$ is a renormalization factor accounting for the phase factor $e^{-2i M \ml{T}}$ from the time evolution of an infinitely heavy particle that generates the Wilson loop.

Furthermore, it turns out that the expectation value of the pure gauge Wilson loop 
\begin{equation} \label{eq:Wloop}
    W[\mathcal{C}] = \frac{1}{N_c} {\rm Tr} P e^{ i g \oint_{\mathcal{C}} {\rm d}s A_\mu(x) \dot{x}^{\mu} } \, ,
\end{equation}
can be described through ``free'' boundary conditions on the $S_5$~\cite{Polchinski:2011im}, i.e., by writing
\begin{align}
\label{eq:nhat-prescription}
    \langle W[\mathcal{C}] \rangle &= \mathcal{N}_{\mathcal{C}} \int \! D\hat{n} \, \langle W_S[\mathcal{C}, \hat{n}] \rangle \,,
\end{align}
where $\mathcal{N}_{\mathcal{C}}$ is a path-dependent (re)normalization factor (the need for it is clear when considering the Euclidean calculation of the heavy quark interaction potential~\cite{Maldacena:1998im}, as the LHS of this equation is bounded by 1 and the RHS isn't [in Euclidean signature]). Equation~\eqref{eq:nhat-prescription} results in Neumann boundary conditions as an equation of motion on the string worldsheet. The fact that Neumann boundary conditions provide a description of a pure gauge Wilson loop was previously argued in~\cite{Alday:2007he}\@.

{\it The Wilson loop that generates the timelike adjoint Wilson line in AdS/CFT.} As described when we formulated the correlator in Eqs.~\eqref{eq:correlator-def} and \eqref{eq:correlator}, an adjoint Wilson line will be generated by performing small deformations on a Wilson loop. Here we discuss the holographic description of this Wilson loop without the deformations (i.e., we set $h^\mu = 0$), solving for the worldsheet that hangs from the closed path $\ml{C}_0$ on the ${\rm AdS}_5$ boundary, where $\ml{C}_0$ consists of two long antiparallel timelike segments of length $\ml{T}$, as introduced earlier. In the large coupling limit $\lambda = R^2/\alpha' \gg 1$, the extrema of the NG action become more and more dominant as $\lambda \to \infty$\@. Thus, it is sufficient to look for the dominant contributions to the path integral in Eq.~\eqref{eq:nhat-prescription} that defines the Wilson loop expectation value:
\begin{equation}
\label{eqn:bound}
    \langle \hat{\mathcal{T}} W[\mathcal{C}_0] \rangle_T = \frac{\mathcal{N}_{\mathcal{C}}}{ \mathcal{Z}} \int \! D\hat{n} \,{\rm Tr}_{\mathcal{H}} \! \left(  e^{-\beta H} \hat{\mathcal{T}} W_S[\mathcal{C}_0,\hat{n}] \right)  \,.
\end{equation}
The dominant contribution to the integral over $S_5$ comes from configurations where $\hat{n}$ takes antipodal positions on opposite sides of the contour, such that $\hat{\ml{T}}W_S[\ml{C}_0,\hat{n}]=1$\@.
This follows from the fact that $e^{-\beta H}$ is a positive definite matrix on the Hilbert space and that the time-ordered Wilson loop $\hat{\mathcal{T}}W_S[\mathcal{C}_0,\hat{n}]$ is constructed from a unitary time-evolution operator, and as such, 
\begin{equation}
    \left| \frac{1}{ \mathcal{Z}} {\rm Tr}_{\mathcal{H}} \! \left(  e^{-\beta H} \hat{\mathcal{T}} W_S[\mathcal{C}_0,\hat{n}] \right) \right| \leq 1 \, .
\end{equation}
An explicit proof of this bound is given in Appendix~\ref{app:W-adj-S5}\@. On the gravity side of the duality, in the limit where $\mathcal{T} \to \infty$, the corresponding extremal surface with minimal energy is the one that hangs from each side of the boundary contours radially into the ${\rm AdS}_5$ bulk.

{\it GGDs in a moving medium.} We consider the case where the rest frame of the plasma is moving with velocity $v$ relative to that of the $Q\bar{Q}$ pair. The holographic setup for a boosted medium is described by the metric of a Lorentz-boosted ${\rm AdS}_5$-Schwarzschild spacetime ($\times S_5$)~\cite{Liu:2006he,Liu:2006nn}.

The calculation of the correlator is analogous to that in a medium at rest~\cite{Nijs:2023dks}, introducing perturbations $h^\mu$ on top of the undeformed contour $\mathcal{C}_0$, except that the string configuration that hangs from each side of $\mathcal{C}_0$ is given by the trailing string of~\cite{Herzog:2006gh,Gubser:2006bz}\@, instead of a string hanging straight into the bulk. This is the lowest energy configuration because $\hat{n}$ takes antipodal positions on opposite sides of the contour. We present this calculation in Appendix~\ref{app:corr-fluct-str}.

Remarkably, we find that the result for the time-ordered correlation function in a moving plasma is equal to that in the static case, but with the substitution $T \to T \sqrt{\cosh{\eta}} = T \sqrt{\gamma}$, where $\gamma = (1 - v^2)^{-1/2}$ is the Lorentz boost factor, in the same way that previous AdS/CFT studies of the heavy quark potential~\cite{Liu:2006he,Liu:2006nn} and diffusion coefficient~\cite{Gubser:2006nz,Casalderrey-Solana:2007ahi} in a medium have observed~\footnote{
A Lorentz transformation of the results for $\kappa_T$, $\kappa_L$ in~\cite{Gubser:2006nz} shows that in the rest frame of the heavy quark one has $\kappa_{T}^{\rm HQ \, rest} = \kappa_{L}^{\rm HQ \, rest} = \pi \sqrt{\lambda} (\sqrt{\gamma} \, T)^3$.}\@. In the rest frame of the heavy quark pair, the longitudinal and transverse components of the chromoelectric field relative to the velocity of the medium are the same. Explicitly, the result is:
\begin{align}
\label{eqn:results}
     [g_{\rm adj}^{\mathcal{T}}]^{\mathcal{N}=4}_{ij}(\omega) = \sqrt{\lambda} T_F \, \delta_{ij} \frac{ (\pi T \sqrt{\gamma})^3 }{12\pi} \! \left( \frac{-i}{F^-_{|\Omega|}(0)} \frac{\partial^3 F^-_{|\Omega|}}{\partial \xi^3}(0) \! \right) \, .
\end{align}
where $F^-_\Omega$ is defined as the regular solution of
\begin{align} \label{eq:F-thermal}
    \frac{\partial^2 F^-_\Omega}{\partial \xi^2} - 2 \left[ \frac{1 + \xi^4}{\xi(1-\xi^4)} - \frac{i \Omega \xi^3}{1-\xi^4} \right] \frac{\partial F^-_\Omega}{\partial \xi} & \\  + \left[ \frac{i \Omega \xi^2}{1-\xi^4} + \frac{\Omega^2 (1 - \xi^6) }{(1-\xi^4)^2} \right] F^-_\Omega & = 0 \nonumber \, .
\end{align}
In the above, $\Omega = \omega/(\pi T \sqrt{\gamma})$, and $\lambda = g^2 N_c$ is the 't Hooft coupling of the $\mathcal{N}=4$ SYM theory. An immediate consequence of Eq.~\eqref{eqn:results} is that the moving medium effect on quarkonium dynamics is that the temperature it experiences gets increased by a factor of $\sqrt{\gamma}$\@. Qualitatively, when the medium is boosted, the light quarks and gluons interacting with quarkonium are more energetic and thus the corresponding quarkonium dynamics occurs faster. Following~\cite{Nijs:2023dks}, we find that the GGD for quarkonium in-medium dynamics in the strong coupling limit is given by
\begin{align}
\label{eqn:gE++}
    [g_{\rm adj}^{++}]^>(\omega) &= 2 \theta(\omega) {\rm Re} \left\{ [g_{\rm adj}^{{\mathcal{T}}}]_{ii}(\omega) \right\} \,.
\end{align}

It is worth emphasizing that from the field theory perspective, the correlation functions that characterize quarkonium and open heavy quark in-medium dynamics are fundamentally different~\footnote{
Comparing to the results of~\cite{Casalderrey-Solana:2006fio}, we find a simple relation between the spectral functions for open heavy quark $\rho_{\rm fund}$ and quarkonium $\rho_{\rm adj}^{++}$ in $\mathcal{N}=4$ SYM:
\begin{equation*}
    \rho_{\rm adj}^{++}(\omega) =  \frac12 \theta(\omega) \big(1 - e^{-\omega/T} \big) \rho_{\rm fund}(\omega) \, .
\end{equation*}
}.
The spectral function for quarkonium $\rho_{\rm adj}^{++}(\omega) = (1 - e^{- \omega/T} ) 
[g_{\rm adj}^{++}]^>(\omega)$ is non-odd in $\omega$ and vanishes at negative frequencies, whereas that for heavy quark diffusion is odd in $\omega$~\cite{Scheihing-Hitschfeld:2023tuz}\@.

{\it Weakly coupled QCD and strongly coupled SYM.}
Finally, we compare the strong coupling and weak coupling results, in order to assess their phenomenological implications and shed light on the physics at intermediate coupling. 
For definiteness, we assume $v=0$ for the purposes of this comparison. The weak coupling result was calculated in~\cite{Scheihing-Hitschfeld:2023tuz,Binder:2021otw}. We provide the explicit expression and details on the choice of renormalization scheme in Appendix~\ref{app:spectral-evaluation}\@.

\begin{figure}
    \centering
    \includegraphics[width=0.48\textwidth]{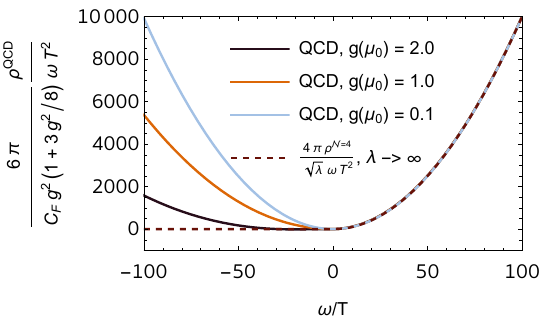}
    \caption{Spectral function for quarkonium transport in weakly coupled QCD with 2 light (massless) quarks for different values of the coupling at the reference scale $\mu_0 \approx 8.1 T$\@. The coupling constant is evolved to high energies using the 2-loop QCD beta function.}
    \label{fig:spectral-2sided}
\end{figure}

Qualitatively, one of the striking features of the result in strongly coupled $\mathcal{N}=4$ SYM is that $[g_{\rm adj}^{++}]^>(\omega<0)=0$\@.
As we can see from Fig.~\ref{fig:spectral-2sided}, increasing the coupling in the perturbative calculation leads to the same feature: if we normalize the spectral function such that its behavior in the ultraviolet (UV) $\omega/T \gg 1$ is fixed, then the spectral function at negative $\omega$ becomes smaller as the coupling increases\@. As such, the trend at weak coupling is compatible with the (supersymmetric) strongly coupled result.

\begin{figure}
    \centering
    \includegraphics[width=0.48\textwidth]{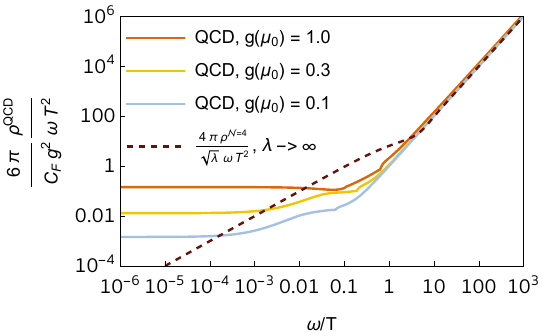}
    \caption{Same as Fig.~\ref{fig:spectral-2sided}, but focusing on the transition between IR and UV physics at positive frequencies. Because the perturbative difference between $\omega > 0$ and $\omega < 0$ is a temperature-independent term, the IR features of the weakly coupled result at negative frequencies are qualitatively the same as those at positive frequencies. The only qualitative difference occurs at $\omega \sim -T$, where the weakly coupled QCD curves cross, in accordance with the $\omega < 0$ behavior of Fig~\ref{fig:spectral-2sided}.}
    \label{fig:spectral-IR}
\end{figure}

We then focus on the infrared (IR) regime $|\omega|/T \lesssim 1$ in Fig.~\ref{fig:spectral-IR}\@.
We have chosen the normalization such that the leading contribution to each curve goes as $\omega^2$ at $\omega/T \gg 1$.
On the one hand, the asymptotic IR behavior of $\rho(\omega)/\omega$ is constant at weak coupling, and linear in $\omega$ at strong coupling.
On the other hand, as before, there is a consistent trend between weak and strong coupling, in the sense that the transition between IR and UV regimes takes place gets pushed to higher values of $\omega/T$ with increasing coupling. This means that, except for the regime $|\omega| \ll T$ (where the convergence of perturbation theory in QCD is generally poor), the perturbative result moves toward the strongly coupled one as the coupling is increased in a consistent trend, both at positive and negative frequencies.

{\it Conclusions.}
We calculated the GGD that characterizes the in-medium dynamics of quarkonium and determines its dissociation and recombination rates in a strongly coupled $\mathcal{N}=4$ SYM plasma moving at velocity $v$ relative to the $Q\bar{Q}$ pair. 
The velocity dependence is a rescaling of the temperature $T$ to $\sqrt{\gamma} \, T$, in consistency with the effect of a ``hot wind'' on quarkonium screening in AdS/CFT~\cite{Liu:2006he,Liu:2006nn}. This effect is not small when the quarkonium momentum is larger than its mass, which is highly relevant for quarkonium production measured in current HIC experiments, and will generically make dissociation/recombination processes faster (as long as the multipole expansion $ \sqrt{\gamma} \, T \ll M v_{\rm rel}$ is under control). This effect will compete with the fact that a $Q\bar{Q}$ pair of higher-$p_T$ generally has less time to interact with the medium.

Furthermore, the strongly coupled result is compatible with the qualitative trends observed by increasing the coupling in perturbative QCD calculations. 
Future phenomenological studies using the GGDs at different couplings have the potential to tell which value of the coupling provides the best description of the experimental data for each quarkonium species. However, our findings imply that this will not be straightforward.
Previous phenomenological studies solved Markovian transport equations such as Boltzmann equations~\cite{Yao:2020eqy} or Lindblad equations~\cite{Brambilla:2022ynh}, in which either the $\omega = -\Delta E < 0$ (if $T \sim \Delta E$) or $\omega = 0$ (if $T \gg \Delta E$) part of $\rho_{\rm adj}^{++}(\omega)$ contributes.
Our results imply that no such contribution exists in the strongly coupled limit, and thus the leading mechanisms driving quarkonium dynamics, coming from the positive $\omega \sim T$ part of $\rho_{\rm adj}^{++}(\omega)$, must be non-Markovian. That is to say, QGP memory effects are not negligible for quarkonium transport in a strongly coupled QGP. Physically, there is no quasi-gluon that a bound $Q\bar{Q}$ pair can absorb resonantly and incoherently in a strongly coupled plasma. Rather, the plasma responds coherently through strong correlations between different points in time, as opposed to behaving as independent, point-like sources.

Our findings motivate formulating the in-medium dynamics of $Q\bar{Q}$ pairs in a general non-Markovian setup, without which it may be impossible to provide reliable phenomenological predictions for quarkonium transport in strongly coupled plasmas. It is also worth exploring at which finite coupling the non-Markovian contribution becomes more important than the Markovian one. In the future, by following this direction, we expect to deepen our knowledge of the QGP's microscopic structure.

\vspace{1cm}

\begin{acknowledgments}
We are grateful for useful comments from Krishna Rajagopal.
This work is supported by the U.S. Department of Energy, Office of Science, Office of Nuclear Physics under grant Contract Number DE-SC0011090. X.Y. is supported by the U.S. Department of Energy, Office of Science, Office of Nuclear Physics, InQubator for Quantum Simulation (IQuS) (https://iqus.uw.edu) under Award Number DOE (NP) Award DE-SC0020970 via the program on Quantum Horizons: QIS Research and Innovation for Nuclear Science.
\end{acknowledgments}

\bibliography{main.bib}

\newpage
\appendix

\begin{widetext}

\section{The timelike adjoint Wilson line in AdS/CFT and the role of the $S_5$} \label{app:W-adj-S5}

In the main text, we claim that
\begin{equation}
    \left| \frac{1}{ \mathcal{Z}} {\rm Tr}_{\mathcal{H}} \! \left(  e^{-\beta H} \hat{\mathcal{T}} W_S[\mathcal{C}_0,\hat{n}] \right) \right| \leq 1 \, .
\end{equation}
for a timelike path $\mathcal{C}_0$ that goes over a straight segment of length $\mathcal{T}$ and then backtracks to its starting point (in what follows, it will be clear that it is not essential for the path to be straight, but it does have to be timelike).

Showing the bound is straightforward once the notation is made explicit. The main ingredient that has to be dealt with carefully is time-ordering. The simplest way to proceed is to define the time-ordered version of the Wilson loop by introducing a more general object that contains it through the differential equation satisfied by the color degrees of freedom of the heavy quarks. Let $\mathcal{W}_{i i_0,j_0 j}$ be such that (in this expression we use the convention that repeated indices are summed; the rest of summations in this section will be made explicit)
\begin{equation} \label{eq:App-def-W-eqn}
    \frac{\rm d}{{\rm d}t} \mathcal{W}_{i i_0,j_0 j} = \left[ i g \left( A_0^a(t) + \hat{n}_1(t) \cdot \phi^a(t) \right) \big[ T^a_{F} \big]_{ii'} \delta_{j'j} - i g \left( A_0^a(t) + \hat{n}_2(t) \cdot \phi^a(t) \right) \big[ T^a_{F} \big]_{j'j} \delta_{ii'} \right] \mathcal{W}_{i' i_0,j_0 j'} \, ,
\end{equation}
with $\mathcal{W}_{i i_0,j_0 j}(t = -\mathcal{T}/2) = \delta_{i i_0} \delta_{j j_0}$. The $S_5$ coordinates are given by $\hat{n}_1(t)$ and $-\hat{n}_2(t)$, representing their values on each side of the contour $\mathcal{C}_0$. The minus sign is necessary to be consistent with the definition~\eqref{eq:Wloop-S}, where there is no sign flip in the prefactor of the scalars caused by the sign flipping of $\dot{x}^{\mu}$. Then, one has $\hat{\mathcal{T}} W_S[\mathcal{C}_0,\hat{n}] = \frac{1}{N_c} \sum_{i,i_0}\mathcal{W}_{i i_0,i_0 i}(t = \mathcal{T}/2)$.

More importantly, $\mathcal{W}$ is a unitary operator on the Hilbert space $\mathcal{H}_{\rm ext} = \mathcal{H} \otimes {\rm Fund}_{N_c} \otimes \overline{\rm Fund}_{N_c}$, which describes the Hilbert spaces of the QGP without any external charge, the heavy quark and the heavy antiquark respectively. As such, we can write
\begin{equation} \label{eq:starting-proof}
    \frac{1}{ \mathcal{Z}} {\rm Tr}_{\mathcal{H}} \! \left(  e^{-\beta H} \hat{\mathcal{T}} W_S[\mathcal{C}_0,\hat{n}] \right) = \frac{1}{\mathcal{Z} N_c} \sum_n e^{-\beta E_n} \sum_{i,j = 1}^{N_c} \langle n , i , i | \mathcal{W} | n, j , j \rangle \, ,
\end{equation}
where we have labeled states in $\mathcal{H}_{\rm ext}$ as $|n, i, j \rangle$, in which $n$ labels the energy eigenstates of $\mathcal{H}$, $i$ labels the color index of ${\rm Fund}_{N_c}$, and $j$ labels the color index of $\overline{\rm Fund}_{N_c}$.
Generally, the action of an operator can be written in terms of its matrix elements. Inserting an identity as a complete set of states yields
\begin{equation}
    \mathcal{W} | n, i , j \rangle = \sum_{m} \sum_{k,l = 1}^{N_c} | m, k, l \rangle [\mathcal{W}]_{mkl,nij}   \, ,
\end{equation}
and the fact that $\mathcal{W}$ is a unitary operator means that we can write its matrix elements in terms of its eigenstates' components $v_{nij}^{(L)}$ as
\begin{equation}
    [\mathcal{W}]_{mkl,nij} = \sum_L v_{mkl}^{(L)*} e^{i \phi_L} v_{nij}^{(L)}
\end{equation} 
where the eigenstates are labelled by $L$.
We then have
\begin{align}
    \frac{1}{ \mathcal{Z}} {\rm Tr}_{\mathcal{H}} \! \left(  e^{-\beta H} \hat{\mathcal{T}} W_S[\mathcal{C}_0,\hat{n}] \right) = \frac{1}{\mathcal{Z} N_c} \sum_n e^{-\beta E_n} \sum_{i,j = 1}^{N_c} [\mathcal{W}]_{nii,njj} &= \frac{1}{\mathcal{Z} N_c} \sum_n e^{-\beta E_n} \sum_{i,j = 1}^{N_c} \sum_L v_{nii}^{(L)*} e^{i \phi_L} v_{njj}^{(L)} \nonumber \\
    &= \frac{1}{\mathcal{Z} N_c} \sum_n \sum_L e^{-\beta E_n} \left| \sum_{i=1}^{N_c} v_{nii}^{(L)} \right|^2 e^{i \phi_L} \, .
\end{align}
Whatever the eigenvectors $v_{nij}^{(L)}$ are, this sum is largest in absolute value if all of the phases $e^{i \phi_L}$ are equal. However, if this is the case, then it follows that $\mathcal{W} = e^{i \phi} \mathbbm{1}$. Therefore, from Eq.~\eqref{eq:starting-proof} we have
\begin{align}
    \left| \frac{1}{ \mathcal{Z}} {\rm Tr}_{\mathcal{H}} \! \left(  e^{-\beta H} \hat{\mathcal{T}} W_S[\mathcal{C}_0,\hat{n}] \right) \right| \leq \frac{1}{\mathcal{Z} N_c} \sum_n e^{-\beta E_n} \sum_{i,j = 1}^{N_c} \langle n , i , i | \mathbbm{1} | n, j , j \rangle &= \frac{1}{\mathcal{Z} N_c} \sum_n e^{-\beta E_n} \sum_{i,j = 1}^{N_c} \delta_{ij} \nonumber \\
    &= \frac{\sum_n e^{-\beta E_n}}{\mathcal{Z}} \frac{\sum_{i= 1}^{N_c} 1}{ N_c} = 1 \, ,
\end{align}
as initially claimed.

Furthermore, this bound is saturated by configurations where $\hat{n}$ takes antipodal positions on the $S_5$. This is easy to see from the defining equation~\eqref{eq:App-def-W-eqn}, because, noting that $\hat{\mathcal{T}} W_S[\mathcal{C}_0,\hat{n}] = \frac{1}{N_c} \mathcal{W}_{i i_0,i_0 i}(t = \mathcal{T}/2)$, it suffices to inspect this differential equation for $i = j$ and $i_0 = j_0$. Explicitly, we have
\begin{align} \label{eq:App-work-W-eqn}
    \frac{\rm d}{{\rm d}t} \mathcal{W}_{i i_0,i_0 i} &= \left[ i g \left( A_0^a(t) + \hat{n}_1(t) \cdot \phi^a(t) \right) \big[ T^a_{F} \big]_{ii'} \delta_{j'i} - i g \left( A_0^a(t) + \hat{n}_2(t) \cdot \phi^a(t) \right) \big[ T^a_{F} \big]_{j'i} \delta_{ii'} \right] \mathcal{W}_{i' i_0,i_0 j'} \nonumber \\
    &= \left[ i g \left( A_0^a(t) + \hat{n}_1(t) \cdot \phi^a(t) \right) \big[ T^a_{F} \big]_{j'i'} - i g \left( A_0^a(t) + \hat{n}_2(t) \cdot \phi^a(t) \right) \big[ T^a_{F} \big]_{j'i'} \right] \mathcal{W}_{i' i_0,i_0 j'} \, ,
\end{align}
which vanishes if $\hat{n}_1 = \hat{n}_2$. As discussed earlier, this corresponds to taking antipodal positions on the $S_5$ for the generalized Wilson loop~\eqref{eq:Wloop-S}. The bound is then saturated because 
\begin{align}
    \frac{\rm d}{{\rm d}t} \mathcal{W}_{i i_0,i_0 i} = 0 \implies \hat{\mathcal{T}} W_S[\mathcal{C}_0,\hat{n}] = \frac{1}{N_c} \mathcal{W}_{i i_0,i_0 i}(t = \mathcal{T}/2) \nonumber = \frac{1}{N_c} \mathcal{W}_{i i_0,i_0 i}(t = -\mathcal{T}/2) = \frac{1}{N_c} \delta_{i i_0} \delta_{i i_0} = 1 \, .
\end{align}
Any other configuration will give a highly oscillatory contribution to the trace over $\mathcal{H}$, and thus its numerical value would be suppressed. Therefore, the dominant contribution indeed comes from the configurations we just described.

One can then also verify on the gravity side of the duality that the extremal worldsheet associated to this configuration is stable and allows for a calculation of the correlator~\eqref{eq:correlator} by solving a set of linear differential equations for the path variations in the dual gravitational description~\cite{Nijs:2023dks}.

As a final comment, we note that the above argument relies crucially on $W_S[\mathcal{C}_0,\hat{n}]$ being constructed from unitary operators. This is true for timelike Wilson loops, but if the path $\mathcal{C}$ is spacelike, then the prefactor $\sqrt{\dot{x}^2}$ of the scalars in the exponential of Eq.~\eqref{eq:Wloop-S} becomes imaginary. Consequently, there is no unitarity bound for such Wilson loops, and thus our preceding argument does not follow through.

\section{Chromoelectric correlator from fluctuations on the trailing string} \label{app:corr-fluct-str}

Here we describe the calculation of the chromoelectric correlator that determines the in-medium dynamics of quarkonium for the case when the QGP is moving with respect to the heavy quark pair. We first discuss the setup of the background worldsheet calculation, which has been studied in the past~\cite{Herzog:2006gh,Gubser:2006bz}, and then proceed to discuss the dynamics of fluctuations on this surface. Some degree of parallel with~\cite{Casalderrey-Solana:2007ahi} will be explicit, but, in the same way as the calculation of~\cite{Nijs:2023dks} differs from that of~\cite{Casalderrey-Solana:2006fio}, there are important conceptual differences to be highlighted.

The reason why the relevant background worldsheet is the string trailing a single heavy quark trajectory is because each side of the contour is located at opposite points on the $S_5$. In the limit $\mathcal{T} \to \infty$ (essentially, $\mathcal{T} \pi T \gg 1$), locally, the lowest energy configuration for the worldsheet hanging from each side of the contour is to fall inwards as if there were only a single Wilson line. Physically, this is consistent with the fact that two heavy quarks in the octet representation cannot form a singlet bound state, and so propagate independently through the same trajectory. This is also consistent with the expectation that the heavy quark potential in the octet channel vanishes in the large $N_c$ limit. Furthermore, this configuration satisfies the expectation $\hat{\mathcal{T}} W[\mathcal{C}_0] = 1$ after subtracting the divergence due to the heavy quark masses.

To characterize this worldsheet, we go to the rest frame of the heavy quarks, where the Wilson lines extend purely along the time direction, and the metric dual to the boosted $\mathcal{N} = 4$ SYM plasma is
\begin{align}
    {\rm d}s^2 = \frac{R^2}{z^2} \bigg\{  - {\rm d}t^2 + {\rm d}x_1^2 + {\rm d}x_2^2 + {\rm d}x_3^2 + \frac{{\rm d}z^2}{f(z)} + z^2 {\rm d}\Omega_5^2  + [1 - f(z)] \left( \cosh^2 \! \eta \, {\rm d}t^2 + \sinh 2\eta \, {\rm d}x_1 {\rm d}t + \sinh^2 \! \eta \, {\rm d}x_1^2 \right) \bigg\} \, ,
\end{align}
where $f = 1 - (\pi T z)^4$. By symmetry considerations, the background worldsheet may be locally described (at times $|t| \ll \mathcal{T}/2$) by
\begin{equation}
    X^{\mu} \to \left( t, \chi(z), 0, 0, z, \hat{n}_0 \right) \, ,
\end{equation}
and the Nambu-Goto action is therefore given by
\begin{equation}
    S_{\rm NG} = - 2 \times \frac{\sqrt{\lambda} \mathcal{T}}{2\pi} \int \frac{{\rm d}z}{z^2} \sqrt{\chi'{}^2 f + \cosh^2 \! \eta - \frac{\sinh^2 \! \eta}{f} } \, .
\end{equation}
(The factor of 2 is due to having two copies of the worldsheet at $\pm \hat{n}_0$.) The extremal surface that solves the equations of motion is determined by
\begin{equation}
    \chi'(z) = - \sinh \eta \frac{\sqrt{1 - f}}{f} = - \sinh \eta \frac{(\pi T z)^2}{1 - (\pi T z)^4} \, ,
\end{equation}
together with $\chi(0) = 0$. One may immediately verify that
\begin{equation}
    \sqrt{\chi'{}^2 f + \cosh^2 \! \eta - \frac{\sinh^2 \! \eta}{f} } = 1 \, ,
\end{equation}
as expected for the mass term that has to be subtracted in order to isolate the expectation value of the Wilson loop. This completely determines the background solution.

Following~\cite{Casalderrey-Solana:2007ahi}, to study the fluctuations on top of this solution it is convenient to introduce a shift in the time coordinate, namely, $\bar{t} = t + F(z)$.
Equivalently, the parametrization of the time coordinate on the worldsheet is now $t = \bar{t} - F(z)$. For obvious reasons, we will drop the bar in what follows. Also, as discussed in~\cite{Nijs:2023dks}, the $i\epsilon$ prescription that enforces time ordering on the Schwinger-Keldysh contour can be accounted for by taking $t$ to be slightly tilted into the negative imaginary direction of the complex time plane. On top of all of these ingredients, we introduce fluctuations parallel and perpendicular to the worldsheet, denoted by $\Delta(t,z), \delta(t,z)$, and $y(t,z)$. As such, the worldsheet parametrization is now
\begin{equation}
    X^{\mu} \to \left( t(1 - i\epsilon) - F(z), \chi(z) + \delta(t,z), y(t,z), 0, z + \Delta(t,z), \hat{n}_0 \right) \, .
\end{equation}
Choosing $F(z)$ to remove cross-terms in the differential equations for the fluctuations lead to choosing it to satisfy
\begin{equation} \label{eq:F-coord-redef}
    F'(z) = \frac{\sinh^2 \! \eta \cosh \eta}{f \cosh^2 \! \eta - \sinh^2 \! \eta} \frac{(1 - f)^{3/2}}{f} \, .
\end{equation}
Concretely, this sets to zero the coefficients of the terms proportional to $y' \dot{y}$ in the quadratic part of the action in the next paragraph (we denote ${\rm d}/{\rm d}z = ()'$, ${\rm d}/{\rm d}t = \dot{()}/(1 - i\epsilon)$). 

From now on, we choose units such that $\pi^2 T^2 \cosh \eta = 1$. (This is allowed because of conformal symmetry.) With this, the Nambu-Goto action, expanded up to quadratic order on $\Delta(t,z), \delta(t,z)$, and $y(t,z)$, reads
\begin{align}
    S_{\rm NG}^{(0-2)} = - \frac{\sqrt{\lambda} (1 - i\epsilon) }{\pi} \!\!  \int  \frac{{\rm d}t \, {\rm d}z}{z^2} & \bigg[ 1  + \frac{z^4 \tanh \eta}{1-z^4} \dot{\delta} + z^2 \tanh \eta \, \delta' + \frac{4 \cosh^2 \! \eta \left( 1 - 5z^4 + 2z^8 + (1 + z^4) \cosh 2\eta \right)}{z (1 - 2z^4 + \cosh 2\eta)^2} \Delta \nonumber \\  
    &   - \frac{2 (1 - z^4) \cosh^2 \! \eta }{1 - 2z^4 + \cosh 2\eta } \Delta'  + \frac{\dot{y}^2}{2 (1 - z^4)}  - \frac{(1 - z^4)}{2} y'{}^2 + \frac{\dot{\delta}^2}{2 (1 - z^4)}  - \frac{(1 - z^4)}{2} \delta'{}^2 \nonumber \\ 
    &  + \frac{z^2 \sinh 2\eta}{1 - 2z^4 + \cosh 2\eta} \left( \frac{1}{ 1 - z^4 } \dot{\delta} \dot{\Delta} + z^2 (\dot{\delta} \Delta' - \delta' \dot{\Delta} ) -  (1 - z^4) \delta' \Delta' \right) \nonumber \\
    &  + \frac{2(1 + 2z^4 + \cosh 2\eta) \sinh 2\eta}{(1 - 2z^4 + \cosh 2\eta)^2} \left( z^3 \dot{\delta} \Delta - z(1-z^4) \delta' \Delta \right) \nonumber \\ 
    &  + \frac{2 z^4 \cosh^2 \! \eta \, \sinh^2 \! \eta}{(1 - 2z^4 + \cosh 2\eta)^2} \left( \frac{1}{1 - z^4} \dot{\Delta}^2 - (1 -z^4) \Delta'{}^2 \right) + \frac{4 z^5 \sinh^2 \! 2\eta }{(1-z^4) (1-2z^4 + \cosh 2\eta)^2} \dot{\Delta} \Delta \nonumber \\
    & + \frac{2 \cosh^2 \! \eta \left( P^{\Delta' \Delta}_{0}(z) + P^{\Delta' \Delta}_{2}(z) \cosh 2\eta + P^{\Delta' \Delta}_{4}(z) \cosh 4\eta \right) }{z(1-2z^4 + \cosh 2\eta)^3} \Delta' \Delta \nonumber \\
    & - \frac{\cosh^2 \! \eta \left( P^{\Delta \Delta}_{0}(z) + P^{\Delta \Delta}_{2}(z) \cosh 2\eta + P^{\Delta \Delta}_{4}(z) \cosh 4\eta + P^{\Delta \Delta}_{6}(z) \cosh 6\eta \right) }{2z^2 (1 - 2z^4 + \cosh 2\eta)^4 } \Delta^2 \bigg] \, , \!
\end{align}
where we have denoted, for brevity,
\begin{align}
    P^{\Delta' \Delta}_{0}(z) &= 3(1-4z^4 + 9z^8 - 4z^{12}) \, , \\ \nn
    P^{\Delta' \Delta}_{2}(z) &= 4(1-3z^4 -z^8 +z^{12}) \, , \\ \nn
    P^{\Delta' \Delta}_{4}(z) &= 1+z^8 \, , \\ \nn
    P^{\Delta \Delta}_{0}(z) &= 30 - 146z^4 + 32z^8 (10 - 17z^4 + 4z^8) \, , \\ \nn
    P^{\Delta \Delta}_{2}(z) &= (45 - 193 z^4 + 290 z^8 + 176 z^{12} - 32z^{16}) \, , \\ \nn
    P^{\Delta \Delta}_{4}(z) &= - 2(-9 + 23z^4 + 8z^8(2 + z^4)) \, , \\ \nn
    P^{\Delta \Delta}_{6}(z) &= (3 + z^4 - 2z^8) \, .
\end{align}

The first thing to note is the presence of linear terms in $\delta$, $\Delta$ in the action. These terms are, naturally, total derivatives, and do not contribute to the equations of motion. However, they could, as written, contribute to the on-shell value of the action. This is not expected nor acceptable on physical grounds, as a non-vanishing contribution at linear order in the perturbations would mean that, firstly, the action was not at an extremum to begin with, and secondly, it would generate a non-vanishing 1-point function of the chromoelectric field on the field theory side of the duality (which is unacceptable because ${\rm Tr} E_i = 0$ where the trace also includes summation over colors). Such considerations imply that consistent solutions for the mode functions of $\delta, \Delta$ will cancel these contributions.

Nonetheless, there is a simpler approach to deal with this potential issue. Geometrically, one can interpret the linear terms in the action for the fluctuations as deformations that are non-orthogonal to the background surface (if they were orthogonal, the action would start at quadratic order). Moreover, the physical perturbations, i.e., those that correspond to a genuine deformation of the surface, are exactly the ones that are orthogonal to the extremal surface. Consequently, the linear terms are associated with the reparametrization invariance of the string worldsheet. Consistently with reparametrization invariance, one can check that the Euler-Lagrange equations derived from extremizing $S_{\rm NG}^{(0-2)}$ with respect to $\delta$ and $\Delta$ are equivalent. Therefore, we can isolate the physical perturbations by setting
\begin{equation} \label{eq:physical-constraint}
    \Delta' = \frac{2}{z} \frac{\cosh^2 \! \eta  - (3 - \cosh^2 \! \eta)z^4 + z^8}{(1 - z^4)(1 - z^4 \, {\rm sech}^2 \eta ) \cosh^2 \! \eta } \Delta + z^2 \tanh \eta \, \frac{1 - z^4 \, {\rm sech}^2 \eta}{1-z^4} \delta' \, .
\end{equation}
This makes the perturbations orthogonal to the worldsheet along $z$. The $y$ perturbations are already orthogonal. As a side note, one may wonder what happens with the $\dot{\delta}$ term, which we have not cancelled by this choice. As it turns out, this can be dealt with in the same way if we had included perturbations for the time component of the worldsheet, i.e., $t(1-i\epsilon) - F(z) \to t(1-i\epsilon) - F(z) + \tau(t,z)$. Including the temporal perturbations $\tau(t,z)$ generates a linear term in the action, which can be chosen to compensate the $\dot{\delta}$ term, thus maintaining the perturbations orthogonal to the worldsheet. One can also verify that the equations of motion for $\tau(t,z)$ are trivial (i.e., all terms in the action that involve this perturbation are total derivatives).

It turns out one can integrate~\eqref{eq:physical-constraint} analytically. Because of local time translation invariance, from here on we Fourier transform $\delta$ and $\Delta$ from functions of time $t$ to functions of frequency $\omega$ (also, whenever we write $\omega$, we actually mean $\omega(1 + i\epsilon)$ due to the slight tilt of the Schwinger-Keldysh contour). The result is
\begin{equation}
    \Delta_\omega(z) = z^2 \tanh \eta \, \frac{1 - z^4 {\rm sech}^2 \eta}{1 - z^4} \left[ \delta_\omega(z) + a_\omega \right] \, ,
\end{equation}
where $a_\omega$ is an integration constant. Then, replacing this constraint in the equation of motion for $\delta$ (or $\Delta$, they are equivalent) to eliminate $\delta$ in favor of $\Delta$, one obtains
\begin{equation}
    \Delta_\omega''(z) - \frac{2}{z} \frac{3 - z^4}{1 - z^4} \Delta_\omega'(z) + \frac{2}{z^2} \frac{5 - z^4}{1 - z^4} \Delta_\omega(z) + \frac{\omega^2}{(1-z^4)^2} \Delta_\omega = \frac{a_\omega z^2 \omega^2 \tanh \eta}{(1 - z^4)^2} \, .
\end{equation}
One may directly verify that the particular solution to this equation is simply $a_\omega z^2 \tanh \eta$. It follows that we can write
\begin{equation}
    \Delta_\omega(z) = a_\omega \tanh \eta \, z^2 + A z^2 \tilde{\Delta}_\omega(z) \, ,
\end{equation}
where $A$ is a normalization constant and $\tilde{\Delta}_\omega(z)$ obeys
\begin{equation}
    \tilde{\Delta}_\omega''(z) - \frac{2}{z} \frac{1 + z^4}{1 - z^4} \tilde{\Delta}_\omega'(z) + \frac{\omega^2}{(1 - z^4)^2} \tilde{\Delta}_\omega(z) = 0 \, .
\end{equation}
Remarkably, this is the same equation that the perturbations satisfy in the case where the direction of the Wilson lines coincide with the rest frame of the medium. The only qualitative difference is the position of the horizon, which here is at $z = (\pi T \sqrt{\cosh \eta})^{-1}$ (where we have temporarily restored units). The solutions to this equation at arbitrary $\omega$ have been studied in~\cite{Nijs:2023dks}. Due to the $i\epsilon$ prescription, the physical, regular solution to the equations of motion is given by only one of the mode functions that solve the homogeneous equation above, which, in the notation of~\cite{Nijs:2023dks}, corresponds to $\tilde{\Delta}_\omega(z) \propto (1 - z^4)^{-i|\omega|/4} F_{|\omega|}^{-}(z)$. Consequently, we have fully determined the mode functions for the fluctuations $\Delta$, $\delta$. Including the normalization constant $A$ for the above mode solutions, we find
\begin{align}
    \delta_\omega(z) &= - \frac{a_\omega \tanh^2 \! \eta \, z^4 }{1 - z^4 \, {\rm sech}^2 \eta} + A \frac{(1 - z^4)^{1 - i |\omega|/4}}{1 - z^4 \, {\rm sech}^2 \eta} F^{-}_{|\omega|}(z) \, , \\
    \Delta_\omega(z) &= a_\omega \tanh \eta \, z^2 + A z^2 (1 - z^4)^{-i |\omega|/4} F_{|\omega|}^{-}(z) \, .
\end{align}
Similarly, the mode functions for the transverse fluctuations $y$ are given by
\begin{equation}
    y_{\omega}(z) = B (1 - z^4)^{-i |\omega|/4} F_{|\omega|}^{-}(z) \, ,
\end{equation}
where $B$ is a normalization constant.

Finally, as discussed in~\cite{Nijs:2023dks}, the time-ordered correlator as a function of $\omega$ is obtained by evaluating the action on the solution with boundary conditions specified by Fourier mode deformations $y(t,z=0), \delta(t,z=0) = e^{- i \omega t}$. Specifically, in position space the correlator is obtained by extracting the quadratic part of the action
\begin{equation}
    [g_{\rm adj}^{\mathcal{T}}]_{ij}(t_2-t_1) = \frac{g^2 T_F}{3 N_c} \langle \hat{\mathcal{T}} E_i^a(t_2) \mathcal{W}^{ab}_{[t_2,t_1]} E_j^b(t_1) \rangle_T = - \frac{i}{12}  \left. \frac{\delta^2  S_{\rm NG}[\mathcal{C};h] }{ \delta h^i(t_2) \delta h^j(t_1)} \right|_{h=0} \, .
\end{equation}
Integrating by parts and using the equations of motion, the on-shell boundary action in the presence of non-vanishing deformations at $z=0$ is given by
\begin{align}
    S_{\rm NG}^{(0-2)} - S_{0} = \frac{\sqrt{\lambda} (1 - i\epsilon) }{\pi}  \int {\rm d}t \lim_{z \to 0} & \bigg[ - \frac{(1 - z^4)}{2 z^2} y y'  - \frac{(1 - z^4)}{2 z^2} \delta \delta' - \frac{\sinh 2\eta \, (1-z^4)}{2(1 - 2z^4 + \cosh 2\eta)} \left( \delta \Delta' + \delta' \Delta \right) \nonumber \\ 
    &  - \frac{ (1-z^4) (1 + 2z^4 + \cosh 2\eta) \sinh 2\eta}{z (1 - 2z^4 + \cosh 2\eta)^2} \delta \Delta  - \frac{2 z^2 (1 -z^4) \cosh^2 \! \eta \, \sinh^2 \! \eta}{(1 - 2z^4 + \cosh 2\eta)^2} \Delta \Delta' \nonumber \\
    & + \frac{\cosh^2 \! \eta \left( P^{\Delta' \Delta}_{0}(z) + P^{\Delta' \Delta}_{2}(z) \cosh 2\eta + P^{\Delta' \Delta}_{4}(z) \cosh 4\eta \right) }{z^3 (1-2z^4 + \cosh 2\eta)^3} \Delta^2 \bigg]  \, ,
\end{align}
where the upper limit of integration $z=1$ gives a vanishing contribution, provided we set $a_\omega = 0$. The reason why the upper limit of integration for the fluctuations is $z=1$ and not $z = \sqrt{\cosh \eta}$ is the following: in the parametrization we have chosen for this calculation, $z=1$ lies on the past infinity hypersurface in the Poincar\'e patch, because the $z$-dependent shift $-F(z)$ in the time coordinate (determined by Eq.~\eqref{eq:F-coord-redef}) goes to $-\infty$ as $z \to 1^-$. This means that the propagation of the perturbations we introduced at the AdS boundary will go outside the Poincar\'e patch when $z>1$, and thus the action for the fluctuations will not receive contributions from $z>1$. (It is important to stress at this point that the $z=1$ contribution to the on-shell value of the action only vanishes if the mode solution is chosen as in~\cite{Nijs:2023dks}, i.e., with the $i\epsilon$ prescription that singles out $F_{|\omega|}^{-}(z)$. The other solutions are discarded because they would give a divergent contribution to the action.)

It would be interesting to study deformations on a Wilson loop of finite extent $\mathcal{T}$, where the way in which the worldsheet is closed at the temporal endpoints must be accounted for explicitly, and see how our current considerations change.

Finally, we are at the point where we can give our result.
Because the mode functions for $\Delta$ go as $z^2$ near $z=0$, the only non-vanishing contributions to $S_{\rm NG}^{(0-2)} - S_0$ come from the $yy'$ and $\delta \delta'$ terms. By analogy with~\cite{Nijs:2023dks}, it follows that
\begin{align}
    [g_{\rm adj}^{\mathcal{T}}]^{\mathcal{N}=4}_{ij}(\omega) =   \frac{\sqrt{\lambda} T_F \delta_{ij}}{ 12\pi} \left( \frac{-i}{F^-_{|\omega|}(0)} \frac{\partial^3 F^-_{|\omega|}}{\partial z^3}(0) \right) \, .
\end{align}
Restoring units by inserting $\pi T \sqrt{\cosh \eta}$ whenever a mass dimension 1 quantity is appropriate recovers the result as announced in the main text.

\section{Evaluation of the spectral function in weakly coupled QCD} \label{app:spectral-evaluation}

The spectral function for quarkonium transport in weakly coupled QCD at positive frequencies was calculated in~\cite{Binder:2021otw}, and its negative frequency part was finally elucidated in~\cite{Scheihing-Hitschfeld:2023tuz}. Up to order $g^4$, it reads
\begin{align}
\label{eqn:rho_UV}
    \rho^{++}_{\rm adj}(\omega)  &=   \frac{g^2 T_F (N_c^2-1) \omega^3 }{ 3  \pi N_c } \nonumber \\ & \times \bigg\{  1  + \frac{g^2}{(2\pi)^2} \bigg[ \left( \frac{11 N_c}{12} - \frac{N_f}{6} \right) \ln \left( \frac{\mu^2}{4 \omega^2} \right)   + N_c \left( \frac{149}{36} - \frac{\pi^2}{6} + \frac{\pi^2}{2} {\rm sgn}(\omega) \right) - \frac{5 N_f}{9}  \bigg]  \nonumber  \\
    & \quad\quad\,\, + \frac{g^2}{(2\pi)^2} \bigg[ \int_0^\infty \!\! \diff k \,  N_f n_F(k) \bigg( -2k \omega + (2k^2 + \omega^2) \ln \left| \frac{k+\omega}{k-\omega} \right|  + 2 k \omega \ln \left| \frac{k^2 - \omega^2}{\omega^2} \right| \bigg) \nonumber \\
    & \quad\quad\quad\quad\quad\,\,\, + \int_{0}^\infty \!\! \diff k \, 2 N_c n_B(k) \bigg( -2 k \omega + (k^2+\omega^2) \ln \left| \frac{k+\omega}{k-\omega} \right|  + k \omega \ln \left| \frac{k^2-\omega^2}{\omega^2} \right| +\mathcal{P} \left( \frac{k^3 \omega}{k^2 - \omega^2} \right)\bigg) \nonumber \\ 
    & \quad\quad\quad\quad\quad\,\,\, + \int_0^\infty \diff k \, \frac{2 N_c n_B(k)}{k} \ml{P} \left( \frac{\omega^2}{ \omega^2 - k^2} \right) \bigg( k^2 \omega + (k^3 + \omega^3) \ln \left| \frac{k-\omega}{\omega} \right| + ( -k^3 + \omega^3) \ln \left| \frac{k + \omega}{\omega} \right| \bigg) \bigg] \bigg\} \nonumber \\
    & + \rho_{\rm HTL}(\omega)
\end{align}
where the hard thermal loop contribution (HTL) can be read off from the heavy quark transport spectral function, as the HTL-resummed diagrams that contribute to them up to $\mathcal{O}(g^4)$ in perturbation theory are the same. Explicitly, it is given by
\begin{align}
    \rho_{\rm HTL}(\omega) =  \frac{g^2 T_F (N_c^2 -1 ) m_D^2 \,\omega}{3\pi N_c} &  \times
 \bigg\{ \int_{\hat\omega}^{\infty} \!  \frac{{\rm d}\hat k \, \hat k}{2} \, 
 \frac{\hat\omega^2
 \Big( 1 - \frac{\hat\omega^2}{\hat k^2}\Big)}
 {
  \Big( 
    \hat k^2 - \hat\omega^2 + \frac12 
     \Big[ 
       \frac{\hat\omega^2}{\hat k^2} + 
       \frac{\hat\omega}{2\hat k} 
       \Big( 1 - \frac{\hat\omega^2}{\hat k^2}\Big) 
       \ln\frac{\hat k + \hat\omega}{\hat k - \hat\omega}
     \Big]
  \Big)^2
 + \Big( 
     \frac{\hat\omega\pi}{4\hat k}
  \Big)^2
       \Bigl( 1 - \frac{\hat\omega^2}{\hat k^2}\Big)^2 
 } \nonumber \\ 
& \quad +  
 \int_{0}^{\infty} \! \frac{ {\rm d}\hat k \, \hat k^3}{2} 
 \bigg[ 
 \frac{
 \theta(\hat k - \hat\omega) 
 }
 {
  \Big( 
    \hat k^2 
     + 1 - 
       \frac{\hat\omega}{2\hat k} 
       \ln\frac{\hat k + \hat\omega}{\hat k - \hat\omega}
       \Big)^2
 + \Big( 
     \frac{\hat\omega\pi}{2\hat k}
  \Big)^2
 } 
 - \frac{1}{(\hat k^2 + 1)^2}
 \bigg]
 \nonumber 
\\ 
& \quad +  
 \left. 
   \frac{ {2 \hat\omega} \hat k_T^3 (\hat\omega^2 - \hat k_T^2)}
   {|3(\hat k_T^2 - \hat\omega^2)^2 -\hat\omega^2|}
 \right|_{\hat k_T^2 - \hat\omega^2 + \frac12
      [\frac{\hat\omega^2}{\hat k_T^2}+
        \frac{\hat\omega}{2\hat k_T} 
       ( 1 - \frac{\hat\omega^2}{\hat k_T^2} ) 
       \ln\frac{\hat\omega + \hat k_T }{\hat\omega - \hat k_T}
       ] \, = \, 0 }
\nonumber 
\\ 
& \quad +  
 \left. 
   \frac{\hat k_E^3 (\hat\omega^2 - \hat k_E^2)}
   { \hat\omega |3(\hat k_E^2 - \hat\omega^2) + 1|}
 \right|_{\hat k_E^2 + 1 -
        \frac{\hat\omega}{2\hat k_E} 
       \ln\frac{\hat\omega + \hat k_E }{\hat\omega - \hat k_E} \, = \, 0 }
  - \frac{\omega^2}{m_D^2} + 
     \frac12  
     \bigg(\ln\frac{2\omega}{m_D} - 1 \bigg)
\bigg\} \,
\end{align}
where, following~\cite{Caron-Huot:2009ncn}, we have denoted $\hat{\omega} = \omega/m_D$, and we have written both the ``naive'' and the ``resummed'' corrections (c.f.~\cite{Burnier:2010rp}) in a single function.

The final step to evaluate this expression is to choose the renormalization scheme, i.e., how to define $\mu$. We choose it following the notion that the best choice of $\mu$ is the one that makes the result the least sensitive to higher order corrections on $g$. In the UV regime, $|\omega| \gg T$, we choose it to compensate for the next-to-leading order (NLO) correction to the $\omega^3$ term of the spectral function. While mathematically we could also choose $\mu$ to compensate the $|\omega|^3$ term, it seems unphysical to let the renormalization group scale depend on the sign of the energy transferred in a physical process (only its magnitude should set the scale). In the IR, we follow~\cite{Burnier:2010rp} and use the electrostatic QCD (EQCD) result of~\cite{Kajantie:1997tt} to set the scale. Putting these together, we choose to interpolate and set the scale for each background temperature $T$ with the following formula:
\begin{equation}
    \mu(\omega, T) = \sqrt{T^2 \exp \left[ \ln(4 \pi) - \gamma_E - \frac{N_c - 8 \ln(2) N_f}{
    2 (11 N_c - 2 N_f)}\right]^2 + 
 \omega^2 \exp \left[\ln(2) + \frac{(6 \pi^2 - 149) N_c + 20 N_f}{6 (11 N_c - 2 N_f)}\right]^2} \, .
\end{equation}
We then choose the value of the coupling constant at the scale $\mu_0$ determined by $\omega = 0$, which means $\mu_0 \approx 8.1 T$, and evolve the coupling constant to higher scales (i.e., $|\omega|>0$) using the 2-loop QCD beta function:
\begin{equation}
    \frac{{\rm d} \alpha_s}{{\rm d} \ln \mu} = -2 \alpha_s \left[ \left(\frac{11 N_c}{3} -\frac{2 N_f}{3} \right) \left(\frac{\alpha_s}{4 \pi}\right) + \left( \frac{34 N_c^2}{3} - 
      \frac{10 N_c N_f}{3} - \frac{(N_c^2 - 1) N_f}{N_c}\right)\left(\frac{\alpha_s}{4 \pi}\right)^2 \right] \, .
\end{equation}

\end{widetext}

\end{document}